\documentclass[epj]{svjour}
\usepackage{graphics}
\usepackage{epsfig}

\begin{document}
\title{ \centerline{ \large BRAG2007 Workshop SUMMARY}  \\ \\
\centerline{The physical meaning of scattering matrix singularities} \\ \centerline{in coupled-channel formalisms}}
\author{ S. Capstick\inst{1}, A. \v{S}varc\inst{2}, L. Tiator\inst{3}, J. Gegelia\inst{3},
M. M. Giannini\inst{4}, E. Santopinto\inst{4}, C. Hanhart\inst{5}, S. Scherer\inst{3}, T.~-S.~H.~Lee\inst{6,7}, T. Sato\inst{7,8} \and N. 
Suzuki\inst{7,8}
}                     
\institute{Department of Physics, Florida State University, Tallahassee, FL 32306-4350, USA \and
Rudjer Bo\v{s}kovi\'{c} Institute, Bijeni\v{c}ka cesta 54, P.O. Box 180, 10002 Zagreb, Croatia\and
Institut f\"{u}r Kernphysik,  Johannes Gutenberg-Universit\"{a}t Mainz, Johann-Joachim-Becher-Weg 45, D-55099 Mainz \and
Dipartimento di Fisica, Universit\`{a} degli Studi di Genova \& I.N.F.N. Sezione di Genova \and
Institut f\"{u}r Kernphysik, Forschungszentrum J\"{u}lich, 52425 J\"{u}lich, Germany \and
Physics Division, Argonne National Laboratory, Argonne, IL 60439 \and
Excited Baryon Analysis Center, Thomas Jefferson National Accelerator Facility, Newport News, Va. 22901 \and
Department of Physics, Osaka University, Toyonaka, Osaka 560-0043, Japan
 }
\date{Received: date / Revised version: date}
\abstract{
 The physical meaning of bare and dressed scattering matrix singularities has been investigated. Special attention has been attributed to the role 
of well known invariance of scattering matrix with respect to the field transformation of the effective Lagrangian. Examples of evaluating bare and 
dressed quantities in various models are given.
} 

\authorrunning{S. Capstick \emph{et el}}
\titlerunning{The physical meaning of scattering matrix singularities ...}
\maketitle

This paper is a collection of different, sometimes conflicting
standpoints presented at the BRAG2007 pre-meeting of the
NSTAR2007 Workshop. It is organized in the following way:\\
The Introduction is written by A. \v{S}varc, Sect. 2.1 by S. Capstick,
2.2 by C. Hanhart, 3.1 by S. Scherer, 3.2 by J. Gegelia, 4.1 by M. Giannini 
and E. Santopinto, 4.2 by  T.~-S.~H.~Lee, T. Sato and N. Suzuki and
the Conclusion by A. \v{S}varc, S. Capstick and L. Tiator.

\section{Introduction}
\label{intro}
Establishing a well defined point of comparison between experimental results and theoretical predictions has for decades been one of the main 
issues in hadron spectroscopy, and the present status is still not satisfactory.  Experiments, via partial wave (PWA) and amplitude analysis (AA), 
can give reliable information on scattering matrix singularities, while quark model calculations usually give information on resonant states 
spectrum in the first order impulse approximation (bare/quenched mass spectrum). And these two quantities are by no means the same. Up to now, in 
the absence of a better recipe, these quantities have usually been directly compared, but the awareness has ripen that the clear distinction 
between the two has to be made. One either has to dress quark-model resonant states spectrum and compare the outcome to the experimental scattering 
matrix poles, or to try to take into account all self-energy contributions which are implicitly included in the measured scattering matrix pole 
parameters, make a model independent undressing procedure and compare the outcome to the impulse approximation quark-model calculations.  The first 
options seems to be feasible but complicated, but the latter one seems to be impossible due to very general field-theory considerations.

We report on investigating both options.

An attempt how to un-quench the constituent quark model of ref. \cite{Cap00}, together with describing all accompanying complications, is 
presented. The procedure seems to be cumbersome, but straightforward.

The second option, undressing the experimentally obtained scattering matrix singularities,  however, seems to be inherently model dependent due to 
very general arguments originating from the local field theory. A simple model, illustrating this claim is presented.

In spite of looking entirely dissimilar, problem of model independent undressing of full scattering matrix singularities seems to be strongly 
correlated to the recent controversy whether the off-shell effects are measurable or not. It is  therefore essential to extend the discussion to 
(un)measurability of off-shell effects as well.

For decades the off-shell properties of two-body amplitudes seemed to be a legitimate measurable quantity, and numerous attempts to get hold of it 
in nucleon-nucleon brems\-strahlung and real and virtual Compton scattering on the nucleon have been made. However, in early 2000-es it became 
apparent that strong field-theoretical arguments do not speak in favor of this claim \cite{Scherer_Fearing:1998wq,Fea00,Sch01}. It seems that it is 
very likely that the well known invariance of scattering matrix with respect to the field transformation of the effective Lagrangian \cite{Gel54} 
makes it possible to transform the off-shell effects into the contact terms for diagrams of the same power counting level. This effectively makes 
the off-shell effects an unmeasurable quantity.

 When applied to the effective two body meson-nucleon amplitudes, this statement implies that the ability of cou\-pled-channel formalisms to 
separate the self-energy term and evaluate the bare scattering matrix poles (singularities in which the meson-exchange effects are fully taken into 
account) is a model dependent procedure. Namely, any method for evaluating self-energy contributions unavoidably demands a definite assumption on 
the analytic form of the off-shell interaction terms, hence introduces model dependent and consequently unmeasurable hadronic shifts.

The invariance of different parameterizations of the scattering matrix singularities with respect to field - redefinitions is also the object of 
our study. Scattering matrix poles are nowadays quantified in two dominant ways: either as Breit-Wigner parameters, i.e. parameters of a 
Breit-Wigner function which is used to locally represent the experimentally obtainable T-matrix, or as scattering matrix poles (either T or K). In 
spite of the fact that it is since Hoehler's  analysis  \cite{Hoe,Hoehler} quite commonly accepted that Breit-Wigner parameters are necessarily 
model dependent quantities, they are still widely used to quantify the scattering matrix poles. Only recently the scattering matrix poles are being 
shown in addition. We demonstrate that within the framework of effective field theory, scattering matrix poles are, contrary to Breit-Wigner 
parameters, unique with respect to arbitrary field-redefinitions.

Bare and dressed scattering matrix quantities have for more then a decade been calculated and presented within a framework of various 
coupled-channel models \cite{Sat,Tia}, and definite correlation between scattering matrix singularities and quark-model quantities has been in 
general established \cite{Int}. However, the most direct connection between full scattering matrix singularities and hadron models with confinement 
forces has been offered in \cite{sl,sl1,yoshimoto} for the various versions of dynamical coupled-channel model. In these models the bare $N^*$ 
states are understood as the excited states of the nucleon if its coupling with the reaction channels is turned off, so the authors naturally 
speculate that the bare $N^*$ states of these models correspond to the predictions from a hadron model with confinement force, such as the 
well-developed constituent quark model with gluon-exchange interactions. The role of hadronic shift model dependence, however,  is not explicitly 
discussed.

A simple conclusion emerges: dressed scattering matrix singularities are the best, model independent meeting point between quark model predictions 
and experiments, and bare quantities in coupled-channel models remain to be legitimate quantities to be extracted only within a framework of a well 
defined model. To understand and interpret them correctly, one has to keep track of the existence of the hadronic mass shifts produced by 
off-shell-ambiguities, and take them fully into account.
\section{Dressing and undressing scattering matrix singularities }
\label{sec:1}
{\subsection{Un-quenching the quark model}
\label{sec:2}
The usual prescription for calculation of the masses of baryons is to
ignore the effects of decay-channel couplings, which is the assumption
that the states are infinitely long-lived. Given that baryon widths
are comparable to the mass splittings between similar states caused by
short-range interactions between the quarks, the effects on baryon
masses of continuum (baryon-meson) states, or equivalently
$qqq$--$q\bar{q}$ components, clearly cannot be ignored. The problem
is that there are many distinct intermediate states which can
contribute substantially to the self energies of baryons through
baryon-meson loops, because of the presence of many thresholds in the
resonance region. Calculations of the effect of two-meson intermediate
states in mesons have been carried out, especially for the interesting
problem of the $\omega$--$\rho$ mass
difference~\cite{GeigerIsgur,PWC} which illustrate the
complications which arise in the case of baryons.
\begin{figure}[h!]
\begin{center}
\epsfig{file=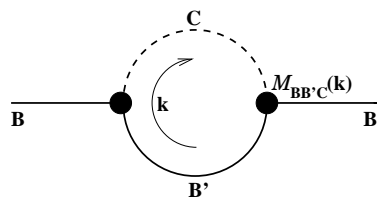, height=2.5cm, angle=0}
\caption{Calculation of baryon self energies in the quark model}
\label{BMloops}
\end{center}
\end{figure}
In order to calculate the self energy of a baryon $B({\bf 0})$ due to
a particular baryon-meson intermediate state $B^\prime(-{\bf k})
C({\bf k})$, as in Fig.~\ref{BMloops}, we require a calculation of
the dependence of the vertex ${\cal M}_{BB^\prime C}(k)$ on the
magnitude $k$ of the loop momentum {\bf k}. This in turn requires a
model of the spectrum (including states not seen in experiment), which
provides wave functions for the baryons, and a model of the $B({\bf
0})\rightarrow B^\prime(-{\bf k}) C({\bf k})$ decay vertices. A
popular choice for the former is some form of constituent quark model,
and for the latter is a pair-creation model such as the $^3P_0$ model
illustrated in Fig.~\ref{3P0}, where baryons decay by the creation of
a quark-antiquark pair with the quantum numbers of the vacuum.
\begin{figure}[t!]
\begin{center}
\epsfig{file=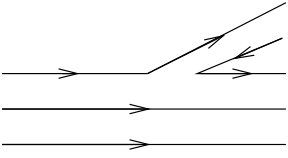, height=2cm, angle=0}
\caption{Pair creation model of baryon decays}
\label{3P0}
\end{center}
\end{figure}

In order to self-consistently calculate the masses of baryons in the
presence of baryon-meson intermediate states, one possible
approach~\cite{Swanson-pc} is as follows. The masses and decays are
calculated using a three-quark Hamiltonian $H_{qqq}$ and a
pair-creation Hamiltonian $H_{\rm pc}$, that depend on strong
coupling, quark mass, and string tension parameters $\alpha_s^0$,
$m_i^0$, and $b^0$, etc., and a pair-creation coupling strength
$\gamma$.  These parameters are usually determined by a fit to the
(dressed) spectrum $E_B$ and decay partial widths in the absence of
Fock-space components higher than $qqq$,
\[
E_B = M_B^0(\alpha_s^0,m_i^0,b^0,...).
\]
The correction due to the loop $B\rightarrow B^\prime C$ is
\[
E_B =  M_B^0(\alpha_s^0,m^0,b^0,...)
+ \Sigma_{B^\prime C}(E_B,E_{B^\prime};\alpha_s^0,m_i^0,b^0),
\]
where
\[
\Sigma_{B^\prime C} = {\cal P}\int d^3k
{
\left\vert \langle B^\prime(-{\bf k})C({\bf k})\vert H_{\rm pc}\vert
B({\bf 0}) \rangle \right\vert^2
\over
E_B - \sqrt{E_{B^\prime}^2+k^2} - \sqrt{E_C^2+k^2} +i\epsilon
}.
\]
The imaginary part of the loop integral is
$\Gamma_{B\rightarrow B^\prime C}/2$. A sum is to be performed
over baryon-meson intermediate states $B^\prime C$ , and the
parameters $\alpha_s^0$, $m_i^0$, $b^0$, and $\gamma$,... are to be
adjusted for self-consistent solution with $E_B$ equal to the dressed
baryon mass. In principle one should solve similar equations for
the meson masses $E_C$.

This procedure is equivalent to second order perturbation theory in
the decay Hamiltonian $H_{\rm pc}$, and allows calculation of the
(momentum-space) continuum $B^\prime C$ component of the dressed
baryon states, and also the mixing $B\rightarrow B^\prime C\rightarrow
B^{\prime\prime}$ between different baryon states caused by the
continuum intermediate states.

Calculation of this kind have been applied to $N$, $\Delta$,
$\Lambda$, $\Sigma$ and $\Sigma^*$ ground and (singly) orbitally
excited states using intermediate states made up of ground state
baryons, with the pseudoscalar mesons $\pi$, $K$, $\eta$,
$\eta^\prime$ in Ref.~\cite{BHM87} and these pseudoscalar
mesons plus the vector mesons $\rho$, $\omega$ and $K^*$ in
Refs.~\cite{Zenczykowski86,S-BG91,Fujiwara93}.
Because there are many baryon-meson thresholds nearby in energy, for
example the $N\rho$ and $\Delta\rho$ thresholds are close to those of
$N(1535)\pi$ or $\Lambda(1405)K$, one should not restrict the
intermediate meson states to $\pi$, or even all pseudoscalars, or the
intermediate baryon states to $N$ and $\Delta$, or even all octet and
decuplet ground states.

Zenczykowski~\cite{Zenczykowski86} showed that if one assumes exact
SU(3)$_f\otimes$SU(2)$_{\rm spin}$ symmetry, that only ground state
ba\-ryons and mesons exist, that all octet and decuplet baryons have
the same mass $M_B^0$ and the same wave function, and that all
pseudoscalar and vector ground-state mesons have the same mass
$M_C^0$ and the same wave function, that all self energy loop
integrals are the same, apart from SU(6)$_{\rm spin-flavor}$ factors
at the vertices. Under these conditions we expect the sum of self
energy contributions to the nucleon and $\Delta(1232)$ masses to be
identical. Interestingly, the sum of the squares of the SU(6)$_{\rm
spin-flavor}$ factors is the same only if we include all
baryon-meson combinations (non-strange, strange, or both) consistent
with the conserved quantum numbers, including both pseudoscalar and
vector mesons. This is true of the self energies of any ground state
baryon, and is also true if the $^3P_0$ model is used to calculate
the vertex factors, as it reduces to SU(6)$_W$ in this limit.

Away from the SU(3)$_f$ limit, Tornqvist and Zen\-czy\-kowski~\cite{TZ84}
were able to show that with the introduction of a simple pattern of
SU(3)$_f$ breaking present in the ground-state baryon and meson mass
spectra, that the usual SU(6) relations for baryon masses are present
in the dressed baryon masses calculated to first order in the symmetry
breaking parameters. This suggests that we can interpret SU(6)
symmetry breaking effects as partly due to spin and flavor-dependent
interactions between the quarks, and partly due to loop effects.

It is clear from this and other calculations that the effects of these
self energies on the spectrum are
substantial. Zenczykowski~\cite{Zenczykowski86} finds many mass
splittings close to those of the dressed pole parameters from analyses,
without spin and flavor-dependent interactions between the
quarks. Other calculations show splittings in the dressed $P$-wave
(lowest orbitally) excited baryons which resemble spin-orbit
effects~\cite{BHM87,S-BG91,Fujiwara93}.
These could cancel against those expected from other sources and
provide a solution to the spin-orbit problem in certain quark models
of baryon masses.

These calculations lack a self-consistent treatment of external and
intermediate baryon states, and so it is not clear that the sum over
intermediate baryon-meson states has converged. Geiger and
Isgur~\cite{GeigerIsgur} demonstrated that this sum does converge, albeit
slowly, using a non-relativistic quark model for baryon masses and
wave functions and a $^3P_0$ model for their decays. Using a covariant
model based on the Schwinger-Dyson Bethe-Salpeter approach was shown
to lead to faster convergence in Ref.~\cite{PWC}. A study of the
dressed masses of $N$ and $\Delta$ ground and $P$-wave excited
baryons~\cite{DM-SC} which involves intermediate pseudoscalar and
vector ground-state mesons and many intermediate baryons (ground
states, and $N$, $\Delta$, $\Lambda$, $\Sigma$, $\Sigma^*$ excited
states up to the second band of negative-parity states at roughly 2100
MeV), representing hundreds of intermediate states, is
underway. Vertex form factors are calculated analytically using mixed
relativized-model~\cite{CI} wave functions and the $^3P_0$
model~\cite{CR1-2}.

This study shows that the usual $^3P_0$ model gives vertices which are
too hard, giving large contributions from high loop momenta. They can
be softened by adopting a pair-creation form factor which decreases as
the relative momentum of the created quark and antiquark
increases. This calculation is currently being reworked to allow
self-consistent renormalization of the quark model parameters. As an
example, in Ref.~\cite{DM-SC} the strong coupling parameter
$\alpha_s^0$ was reduced in order to take into account the additional
$\Delta-N$ splitting in the sums over baryon-meson loops contributing
to the self energies of both of these states. Similarly, in their
calculation of these effects in mesons, Geiger and
Isgur~\cite{GeigerIsgur} showed that the formation of an intermediate
meson pair was equivalent to string breaking, which has the effect of
renormalizing the meson string tension. Barnes and
Swanson~\cite{TednEric} have examined shifts in the charmonium
spectrum due to $D$, $D^*$, $D_s$ and $D_s^*$ meson pairs.

In conclusion, the next Fock-space component is likely more important
than differences among $qqq$ models. Calculating its effects requires
the use of a full set of SU(6)-related intermediate states, spatially
excited intermediate baryons, and a careful treatment of mixing
effects. Renormalization of the parameters in the quark model
parameters such as $\alpha_s$, the quark masses, the string tension,
and the pair-creation strength needs to be carried out
systematically. This requires examining the mass shifts of more than
just $N$, $\Delta$ and their negative-parity excitations. Decay
vertices need additional suppression when the dressed masses of
external states are well above the threshold for an intermediate
state, which is the case in relativistic models.

\subsection{Undressing the dressed scattering matrix singularities}
\label{sec:3}
To get a better understanding of the relation of bare
quantities to dressed quantities it is sufficient
to study a system of two nucleon like states ($N$ and $R$)
coupled to a scalar field ($\sigma$)~\cite{collab}. A possible
Lagrangian reads
\begin{eqnarray}
\nonumber
{\cal L}_1=\bar N\left(i\partial \!\!\! / -M_N\right)N+\bar R\left(i\partial
  \!\!\! / -M_R^{0}\right)R
\\
+\frac12\left(\partial^\mu\sigma\partial_\mu \sigma -m^2\right)
+g\sigma\left(\bar R N+\bar NR\right) ... \ .
\label{toyI}
\end{eqnarray}
Here the superscript $0$ indicates that masses are bare quantities
that undergo dressing beyond tree level\footnote{In principle also the
  coupling $g$ and the mass of $\sigma$ and $N$ are bare quantities, however,
  to ease notation we drop the corresponding superscript, for in what follows
we focus solely on the self energy of the $R$ field.}.  The resulting
vertex is shown as diagram (a) in Fig. \ref{vert}.  The dots indicate possible
more complex terms, like contact terms of the type $\sigma^2\bar RR$ (see Fig.
\ref{vert}(c)). However, in phenomenological studies those are rarely
included. From this Lagrangian we may now calculate observables like
scattering amplitudes. To keep things simple we focus only on the self--energy
of the $R$ field. The corresponding diagram is shown in Fig.
 \ref{selfen}(a). The real part of this diagram provides the so called
hadronic shift --- the difference between the bare mass and the physical mass
--- and the imaginary part the width.
\begin{figure}[t!]
\begin{center}
\epsfig{file=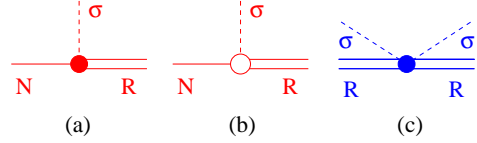, height=2cm, angle=0}
\caption{Vertices from the interaction Lagrangians of Eqs.
 (\ref{toyI}) and (\ref{toyII}).}
\label{vert}
\end{center}
\end{figure}

A theorem based on very general assumptions in field theory states
that {\it if two fields $\phi$ and $\chi$ are related non--linearly
($\phi = \chi F(\chi)$ with $F(0)=1$)
then the same observables arise if one calculates
with $\phi$ using ${\cal L}(\phi)$ or with $\chi$
using  ${\cal L}(\chi F(\chi))$}~\cite{fieldindep}.
\begin{figure*}[t!]
\begin{center}
\epsfig{file=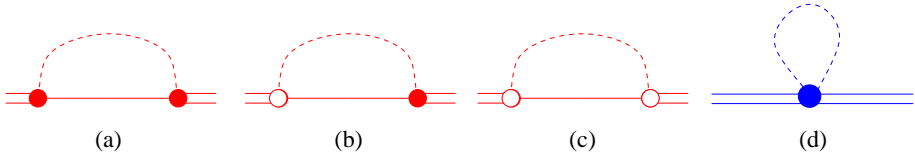, height=2cm, angle=0}
\caption{Selfenergies for the $R$ field to one
loop order from toy--model I (Eq.
 (\ref{toyI})) and toy--model II (Eq. \ref{toyII}).}
\label{selfen}
\end{center}
\end{figure*}
Thus instead of the fields in Eq. (\ref{toyI}) we
may switch to a modified nucleon field defined through
$$
N \longrightarrow N'=N+\alpha \sigma R \ ,
$$
where $\alpha$ is an arbitrary, real parameter. Expressed
in terms of $N'$ the interaction part of the original Lagrangian now reads
\begin{eqnarray} \nonumber
{\cal L}_{1'}^I
&=&g\sigma\left(\bar R N'+\bar N'R\right)
                         -\alpha \sigma\left(\bar R\left(i\partial \!\!\! /
-M_N\right)N' + \mbox{h.c.} \right)\\
 &+&\sigma \bar R\left[\alpha^2\left(i\partial \!\!\! / -M_N\right)-2g\alpha\right]R
\sigma + ... \ .
\label{1prime}
\end{eqnarray}
In addition to the vertex
of the previous Lagrangian now two new structures appear: a
momentum dependent $RN\sigma$ vertex (depicted in Fig. \ref{vert}(b))
and a $\sigma^2\bar RR$ vertex (depicted in Fig. \ref{vert}(c)).
The resulting contributions to the $R$--selfenergy from
this Lagrangian are shown in
Fig. \ref{selfen}(a)--(d). The field theoretic theorem quoted above
gives that the total self energy contribution from the modified
Lagrangian is identical to that of the original Lagrangian with the
same parameters.  Especially, the hadronic shift remains the same and
it seems that indeed it is a well defined quantity. However,
the problem is that we do not know the $true$ hadronic Lagrangian.
Thus, starting from Eq. (\ref{toyI}) is as justified as starting
from the following interaction Lagrangian
\begin{eqnarray} \nonumber
{\cal L}_{2}^I&=&g\sigma\left(\bar R N+\bar NR\right)
                         -\alpha \sigma\left(\bar R\left(i\partial \!\!\! /
-M_N\right)N + \mbox{h.c.} \right)\\ & & \qquad\qquad\qquad\qquad\qquad
+  \ ... \ .
\label{toyII}
\end{eqnarray}
Obviously the only difference to the previous equation is that
the $\sigma^2\bar RR$ vertex was abandoned. On the one loop level
thus the only difference compared to the previous expression
for the $R$--selfenergy is that tadpole diagrams were removed.
Since this class of diagrams does not lead to non--analyticities,
their effect can always be absorbed into the bare mass
and the wave function renormalization of the $R$ field.
Therefore, with properly adjusted parameters, the self--energy
is the same to one--loop between the theory that follows from
${\cal L}_1$ and that from ${\cal L}_2$.

Is there any way in practise to decide, which one of the two
Lagrangians is to be preferred? The answer to this question is $no$
for the following reasons: although the $\sigma^2\bar RR$ contact
term can contribute to the $R$ self energy at three loop order, this
is of no practical significance, since not only has any effective
Lagrangian a too limited range of applicability and accuracy to
allow for the extraction of such effects but also a complete
treatment should include anyway direct $\sigma R\to \sigma R$
transitions in both Lagrangians in addition to the terms given
explicitly above. The latter argument also applies to information
deduced from $\sigma R\to \sigma R$ cross sections. Therefore there
is in practice no way to decide which one of the two interaction
Lagrangians --- Eq. (\ref{toyI}) or Eq.  (\ref{toyII}) --- is to be
preferred. As outlined above, however, quantities like the
self--energies of the resonance $R$ are different in the two
approaches and consequently the bare masses as extracted from fits
to experiment are different. We therefore conclude that bare masses
(or in general bare quantities) do not have any physical
significance.

The question studied here is very closely linked to the question of
measurability of off--shell effects. The argument just presented can also be
used as yet another illustration that off--shell effects are not observable.
This already follows from a comparison of Eq.~(\ref{toyI}) and
Eq.~(\ref{1prime}). As argued above both lead to identical observables.
Especially, the on--shell $RN\sigma$ vertex that can be related to the decay
width from $R\to \sigma N$ is the same for both models.  However, in our
example for off--shell nucleons the vertex can be anything.  In general,
within a consistent field theory off--shell effects either can be absorbed
into counter--terms or have to cancel exactly.  The same issue is discussed
for bremsstrahlung in Ref.~\cite{Sch01}.  Another illustrative example of
the cancellation of off--shell effects is provided in Ref.~\cite{pp2dpi} for
the reaction $NN\to NN\pi$.

It should be stressed that the question in focus here is very
different to that of the relation between two--nucleon and
three--nucleon observables and the presence of three--body forces.
The main difference is that in the few nucleon systems it is
possible to construct three--body forces that are consistent with
the two--nucleon interaction used, e.g. within effective field
theory --- for a recent review see Ref.~\cite{evgenirev}. Changing
the two--nucleon interactions leads also to controlled changes in
the three--body forces in the sense sketched above.  However, what
would be needed for a model independent extraction of bare hadron
masses would be a method to identify $the$ hadronic interaction that
is the one that matches to the particular quark model, thus a
connection is needed between two systems with very different degrees
of freedom. We argue that it follows from the reasoning above that
this identification can not be made as a matter of principle.
However, the inclusion of hadronic loops within the quark model, as
sketched, e.g.  in the presentation by Simon Capstick, is obviously
justified.

We therefore have to conclude that the only quantities relevant for
spectroscopy that can be extracted
from experiment are resonance poles and the corresponding residues.
However, this is still a lot for both quantities contain important
structure information like the amount of $SU(3)$ violation or
even the very nature of the state~\cite{evidence}.
A extraction of poles and residues from the data needs
coupled--channel codes of the type of Refs.~\cite{krehl,lee,svarc}
with the correct analytical properties and consistent with
unitarity. Only the then a controlled analytic continuation
to the complex plain is possible.

\section{Field-theory considerations}
\subsection{From off-shell to on-shell kinematics}
 \label{sec:4}
  It is a natural and
legitimate question to ask whether the off-shell behavior of
particular interaction vertices is unique and whether it is possible
to extract such behavior from empirical information similarly as
one, say, extracts the electromagnetic form factors of the nucleon
from elastic electron scattering.
   In this context one might think of the electromagnetic interaction of
a bound (off-shell) nucleon or the investigation of the off-shell
nucleon-nucleon amplitude entering the
nucleon-nucleon-bremsstrahlung \linebreak process.
   For the case of pions, Compton scattering \cite{Scherer_Scherer:1994aq}
and pion-pion bremsstrahlung \cite{Scherer_Fearing:1998wq} were
discussed using chiral perturbation theory (ChPT) at lowest order.
   It was shown that off-shell effects with respect to the effective
pion fields depend on both the model used and the choice of
representation for the fields.
   From that the conclusion was drawn that off-shell effects are not only
model dependent but also representation dependent, making a unique
extraction of off-shell effects impossible.
   The spin-1/2 case was discussed in Ref.\ \cite{Fea00}.

  A related situation occurs when one is interested in corrections
to current-algebra results obtained from the partially conserved
axial-vector current (PCAC) relation
\begin{equation}
\label{Scherer_PCAC} \partial_\mu A^{\mu,a}= M_\pi^2 F_\pi \Phi^a,
\end{equation}
   where $A^{\mu,a}$ is the isovector axial-vector current
and $\Phi^a$ is a renormalized field operator creating and
destroying pions;
   $M_\pi$ and $F_\pi$ denote the pion mass and decay constant,
   respectively.
   While predictions of current algebra and the PCAC relation involve
the so-called soft-pion limit, \linebreak $\lim_{q_0\to 0}
\lim_{\vec q\to 0}[\cdots]$, amplitudes for physical pions are to be
taken at $q^2=M_\pi^2$.
   Can the connection between soft-pion kinematics
and on-shell kinematics be uniquely determined?
   The answer is yes, if the problem is entirely
formulated in terms of the relevant QCD Green functions.

   We will illustrate these issues in the framework of
ChPT \cite{Gasser:1983yg} which establishes a systematic
connection with the underlying field theory, namely, QCD.
   Let us first discuss off-shell effects with
respect to the effective fields.
   To that end, we consider $\pi\pi$ scattering
at lowest order in ChPT (see Sect.\ 4.6.2 of Ref.\
\cite{Scherer_Scherer:2002tk} for more details):
$${\cal L}_2=\frac{F^2_\pi}{4}\mbox{Tr}\left[\partial_\mu U (\partial^\mu U)^\dagger\right]
+\frac{F^2_\pi M_\pi^2}{4}\mbox{Tr}(U+U^\dagger),$$ where $M^2_\pi=2
B \hat{m}$. $B$ is related to the quark condensate $\langle \bar{q}
q\rangle_0$ in the chiral limit and $\hat m$ is the average of the
$u$- and $d$-quark masses \cite{Gasser:1983yg}; $U$ is an
SU(2) matrix containing the pion fields. We will use two alternative
parameterizations of $U$:
\begin{eqnarray*}
U(x)&=&\frac{1}{F_\pi}\left[\sigma(x) 
+i\vec{\tau}\cdot\vec{\pi}(x) \right],\quad
\sigma(x)=\sqrt{F^2_\pi-\vec{\pi}^2(x)},\\
U(x)&=&\exp\left[i\frac{\vec{\tau}\cdot\vec{\phi}(x)}{F_\pi}\right]
=\cos\left(\frac{\phi}{F_\pi}\right)
+i\vec{\tau}\cdot\hat{\phi} \sin\left(\frac{\phi}{F_\pi}\right).
\end{eqnarray*}
The $\sigma$ and exponential parameterizations are related by a
field transformation (change of variables)
$$
\frac{\vec{\pi}}{F_\pi}=\hat{\phi}\sin\left(\frac{\phi}{F_\pi}\right)
=\frac{\vec{\phi}}{F_\pi}\left(1-\frac{1}{6}\frac{\vec{\phi}^2}{F^2_\pi}
+\cdots\right).
$$
   The relevant $\pi\pi$ interaction Lagrangians read
\begin{eqnarray*}
{\cal L}_2^{4\pi}&=&\frac{1}{2F^2_\pi}\partial_\mu
\vec{\pi}\cdot\vec{\pi}
\partial^\mu \vec{\pi}\cdot\vec{\pi}
-\frac{M_\pi^2}{8 F^2_\pi}(\vec{\pi}^2)^2,\\
{\cal L}_2^{4\phi}&=&\frac{1}{6F^2_\pi}(\partial_\mu
\vec{\phi}\cdot\vec{\phi}
\partial^\mu \vec{\phi}\cdot\vec{\phi}-\vec{\phi}^2 \partial_\mu\vec{\phi}
\cdot \partial^\mu\vec{\phi}) +\frac{M_\pi^2}{24
F^2_\pi}(\vec{\phi}^2)^2.
\end{eqnarray*}
   Observe that the two interaction Lagrangians depend differently on the
respective pion fields.
   For Cartesian isospin indices $a,b,c,d$ the Feynman rules for
the scattering process
$\pi^a(p_a)+\pi^b(p_b)\to\pi^c(p_c)+\pi^d(p_d)$ read, respectively,
\begin{eqnarray*}
{\cal
M}_2^{4\pi}&=&i\left(\delta^{ab}\delta^{cd}\frac{s\!-\!M^2_\pi}{F^2_\pi}\!
+\!\delta^{ac}\delta^{bd}\frac{t\!-\!M^2_\pi}{F^2_\pi}
\!+\!\delta^{ad}\delta^{bc}\frac{u\!-\!M^2_\pi}{F^2_\pi}\right),\\
\label{4:6:mpipi2}{\cal M}_2^{4\phi}&=&{\cal
M}_2^{4\pi}\nonumber\\
&&\hspace{-2em}-\frac{i}{3F_\pi^2}
\left(\delta^{ab}\delta^{cd}+\delta^{ac}\delta^{bd}+\delta^{ad}\delta^{bc}
\right) \left(\Lambda_a+\Lambda_b+\Lambda_c+\Lambda_d\right),
\end{eqnarray*}
where we introduced $\Lambda_k=p_k^2-M^2_\pi$ and the usual
Mandelstam variables $s=(p_a+p_b)^2$, $t=(p_a-p_c)^2$, and
$u=(p_a-p_d)^2$ satisfying $s+t+u=p_a^2+p_b^2+p_c^2+p_d^2$.
  If the initial and final pions are all on the mass shell,
i.e., $\Lambda_k=0$, the scattering amplitudes are the same, in
agreement with the equivalence theorem of field theory
\cite{Scherer_ETFT}. On the other hand, if one of the momenta of the
external lines is off mass shell, the amplitudes ${{\cal
M}_2^{4\pi}}$ and ${{\cal M}_2^{4\phi}}$ differ.
   In other words, a ``direct'' calculation of
$\pi\pi$ scattering in terms of the effective fields gives a unique
result independent of the parameterization of $U$ only for the
on-shell matrix element.

   According to the standard argument in nucleon-nucleon bremsstrahlung
one would now try to discriminate between different on-shell
equivalent $\pi\pi$ amplitudes through an investigation of the
reaction $\pi^a(p_a)+\pi^b(p_b)\to\pi^c(p_c)+\pi^d(p_d)+\gamma(k)$.
   This claim was critically examined in Refs.\
\cite{Scherer_Fearing:1998wq,Fea00}.
   To that end the electromagnetic field is included through the
   covariant derivative $D_\mu U=\partial_\mu U +ieA_\mu[Q,U]$
   where $Q=\mbox{diag}(2/3,-1/3)$ is the quark charge matrix.
  In the $\sigma$ parameterization, the total bremsstrahlung amplitude
is given by the sum of only such diagrams, where the photon is
radiated off the initial and final charged pions, respectively.
  One may then ask how the different off-shell behavior of
the $\pi\pi$ amplitude of ${{\cal M}_2^{4\phi}}$ enters into the
calculation of the bremsstrahlung amplitude.
   Observe, in this context, that the exponential
parameterization generates electromagnetic interactions involving
$2n$ pion fields, where $n$ is a positive integer.
   In the exponential parameterization an additional
$4\phi\gamma$ interaction term relevant to the bremsstrahlung
process is generated.
  Hence the total tree-level amplitude now contains an additional
four-pion-one-photon contact diagram.
    Combining the contribution due to the off-shell behavior in
the $\pi\pi$ amplitude ${\cal M}^{4\phi}_2$ with the contact-term
contribution, we found a precise cancelation of off-shell effects
and contact interaction such that the final results are the same for
both parameterizations.
   This is once again a manifestation of the equivalence theorem
\cite{Scherer_ETFT}.
   What is even more important in the present context is the observation
that the two mechanisms, i.e.\ contact term vs.\ off-shell effects,
are indistinguishable since they lead to the same measurable
amplitude.

   Now, what about the off-shell behavior of QCD Green functions?
   The method developed by Gasser and Leutwyler \cite{Gasser:1983yg}
deals with Green functions of color-neutral, Hermitian quadratic
forms involving the light-quark fields $q=(u,d)^T$ of QCD and their
interrelations as expressed in the Ward identities.
   In particular, these Green functions can, in principle, be calculated
for any value of squared momenta even though ChPT is set up only for
a low-energy description.
   For the discussion of $\pi\pi$ scattering one considers the four-point
function \cite{Scherer_Gasser:1983yg}
\begin{equation}
\label{4:6:fpfpppp} G_{PPPP}^{abcd}(x_a,x_b,x_c,x_d)\equiv \langle
0|T[P^a(x_a)
\cdots P^d(x_d)]| 0\rangle
\end{equation}
with the pseudoscalar quark density $P^a=i\bar{q}\gamma_5 \tau^a q$.
   In order so see that Eq.\ (\ref{4:6:fpfpppp}) can indeed be related to
$\pi\pi$ scattering, we investigate the matrix element of $P^a$
evaluated between a single-pion state and the vacuum
\cite{Scherer_Gasser:1983yg}:
\begin{equation}
\label{4:6:pipiv} \langle 0|P^a(0)|\pi^b(q)\rangle
\equiv\delta^{ab}G_\pi.
\end{equation}
   The coupling of an external pseudoscalar source $p$ to the Goldstone
bosons is given by
\begin{eqnarray}
\label{4:6:l2ext2} {\cal L}_{\rm ext}
&=&i\frac{F_\pi^2B}{2}\mbox{Tr}(pU^\dagger-Up)\nonumber\\
&=& \left\{\begin{array}{l}
2B F_\pi p^a \pi^a,\\
2B F \pi p^a\phi^a[1-\vec{\phi}\,^2/(6F_\pi^2)+\cdots],
\end{array}\right.
\end{eqnarray}
where the first and second lines refer to the $\sigma$ and
exponential parameterizations of $U$, respectively.
   From Eq.\ (\ref{4:6:l2ext2}) we obtain $G_\pi = 2B F_\pi$ independent of
the parameterization used which, since the pion is on-shell, is a
consequence of the equivalence theorem \cite{Scherer_ETFT}.
   As a consistency check, let us verify the PCAC relation
from the QCD Lagrangian
\begin{displaymath}
\partial_\mu A^{\mu,a}=\hat{m}i\bar{q}\gamma_5 \tau^a
q\equiv\hat{m}P^a,
\end{displaymath}
evaluated between a single-pion state and the vacuum.
   The axial-vector current matrix element obtained from
   ${\cal L}_2$ reads
\begin{equation}
\label{4:6:axialcurrentpion} \langle 0|A^{\mu,a}(x)|\pi^b(q)\rangle
= i q^\mu F_\pi e^{-iq\cdot x}\delta^{ab}.
\end{equation}
   Taking the divergence implies $M_\pi^2 F_\pi=\hat m G_\pi$.
In other words,
\begin{equation}
\label{scherer:pionfield}
 \Phi^a(x)\equiv \frac{\hat m P^a(x)}{M_\pi^2 F_\pi}
\end{equation}
can serve as a so-called {\em interpolating} pion field in the LSZ
reduction formula.
   Using Eq.~(\ref{scherer:pionfield}), the
reduction formula relates the $S$-matrix element of $\pi\pi$
scattering to the QCD Green function involving four pseudoscalar
densities
\begin{eqnarray*}
\lefteqn{S_{fi}=
\left(\frac{-i}{G_\pi}\right)^4(p_a^2-M_\pi^2)\cdots(p_d^2-M_\pi^2)}\\
&&\!\!\times\!\int d^4 x_a \cdots d^4 x_d
 \,e^{-i p_a \cdot x_a} \cdots e^{i p_d \cdot x_d}
G_{PPPP}^{abcd}(x_a,x_b,x_c,x_d).
\end{eqnarray*}
   Using translational invariance, let us define the momentum space
Green function as
\begin{eqnarray*}
\label{4:6:msgf} \lefteqn{ (2\pi)^4
\delta^4(p_a+p_b+p_c+p_d)F^{abcd}_{PPPP}(p_a,p_b,p_c,p_d)=}\nonumber
\\
&& \int d^4 x_a d^4 x_b d^4 x_c d^4 x_d\,  e^{-i p_a \cdot x_a}
e^{-ip_b\cdot x_b} e^{-i p_c\cdot x_c} e^{-i p_d x_d}\nonumber\\
&&\times G_{PPPP}^{abcd}(x_a,x_b,x_c,x_d),\nonumber
\end{eqnarray*}
   where we define all momenta as incoming.
   The usual relation between the $S$ matrix and the $T$ matrix,
$S=I+iT$, implies for the $T$-matrix element $\langle f|T|i\rangle=
(2\pi)^4\delta^4(P_f-P_i){\cal T}_{fi}$ and, finally, for ${\cal
M}=i{\cal T}_{fi}$:
\begin{equation}
\label{4:6:calMlsz} {\cal M}= \frac{1}{G_\pi^4}
\left[\prod_{k=a,b,c,d}\lim_{p_k^2\to M_\pi^2}
(p_k^2-M_\pi^2)\right] F^{abcd}_{PPPP}
\end{equation}
with $ F^{abcd}_{PPPP}\equiv F^{abcd}_{PPPP}(p_a,p_b,-p_c,-p_d)$. We
will now determine the Green function $ F^{abcd}_{PPPP}$ using the
$\sigma$ and exponential parameterizations for $U$.
   In the first
parameterization we only obtain a linear coupling between the
external pseudoscalar field and the pion field [see Eq.\
(\ref{4:6:l2ext2})] so that only one Feynman diagram contributes
\begin{equation}
\label{4:6:fabcd1} F^{abcd}_{PPPP}=(2B F_\pi)^4
\frac{i}{p_a^2-M_\pi^2}\cdots\frac{i}{p_d^2-M_\pi^2} {\cal
M}^{4\pi}_2.
\end{equation}
   The Green function $F^{abcd}_{PPPP}$ depends on six independent Lorentz
scalars which can be chosen as the squared invariant momenta $p^2_k$
and the three Mandelstam variables $s$, $t$, and $u$ satisfying the
constraint $s+t+u=\sum_k p_k^2$.

   Using the second parameterization we will obtain a contribution
which is of the same form as Eq.~(\ref{4:6:fabcd1}) but with ${\cal
M}^{4\pi}_2$ replaced by ${\cal M}^{4\phi}_2$.
   Clearly, this is not yet the same result as Eq.\ (\ref{4:6:fabcd1})
because of the terms proportional to $\Lambda_k$ in ${\cal
M}^{4\phi}_2$.
   However, in this parameterization the external pseudoscalar field
also couples to three pion fields [see Eq.\ (\ref{4:6:l2ext2})],
resulting in four additional contributions
\begin{displaymath}
\Delta_a F^{abcd}_{PPPP} +\cdots +\Delta_d F^{abcd}_{PPPP}
\end{displaymath}
with
\begin{eqnarray}
\label{4:6:deltafabcd2}
\lefteqn{\Delta_a F^{abcd}_{PPPP}(p_a,p_b,-p_c,-p_d)}\nonumber\\
&=&(2B F_\pi)^4
\frac{i}{p_a^2-M_\pi^2}\cdots\frac{i}{p_d^2-M_\pi^2}\nonumber\\
&&\times \frac{i\Lambda_a}{3 F_\pi^2}(\delta^{ab}\delta^{cd}
+\delta^{ac}\delta^{bd}+\delta^{ad}\delta^{bc}),
\end{eqnarray}
and analogous expressions for the remaining $\Delta F$s.
   In total, we find a complete cancelation with those terms proportional to
$\Lambda_k$ (in the second parameterization) and the end result is
identical with Eq.\ (\ref{4:6:fabcd1})!
   Finally, using $G_\pi=2 B F_\pi$ and inserting the result of
Eq.\ (\ref{4:6:fabcd1}) into Eq.\ (\ref{4:6:calMlsz}) we obtain the
same scattering amplitude as in the ``direct'' calculation of ${\cal
M}^{4\pi}_2$ and $ {\cal M}^{4\phi}_2$ evaluated for on-shell pions.

   This example serves as an illustration that the method of Gasser and
Leutwyler generates unique results for the Green functions of QCD
for arbitrary four-momenta.
   There is no ambiguity resulting from the choice of variables used
to parameterize the matrix $U$ in the effective Lagrangian.
   These Green functions can be evaluated for arbitrary (but small)
four-momenta.
   Using the reduction formalism, on-shell matrix elements such as the
$\pi\pi$ scattering amplitude can be calculated from the QCD Green
functions.
   The result for the $\pi\pi$ scattering amplitude as derived from
Eq.\ (\ref{4:6:calMlsz}) agrees with the ``direct'' calculation of
the on-shell matrix elements of ${\cal M}^{4\pi}_2$ and $ {\cal
M}^{4\phi}_2$.
   On the other hand, the Feynman rules of
${\cal M}^{4\pi}_2$ and $ {\cal M}^{4\phi}_2$
when taken {\em off shell}, have to be
considered as intermediate building blocks only and thus need not be
unique.

\subsection{Model (in)dependence of pole positions and Breit-Wigner parameters}
 \label{sec:5}
   A popular definition of masses of unstable particles corresponding to a
(relativistic) Breit-Wigner formula makes use of the zero of the
real part of the inverse propagator.
   It has been shown that such a definition leads to
   field-redefinition and gauge-parameter dependence of the mass
starting at two-loop order
\cite{Willenbrock:1990et,Valencia:1990jp,Sirlin:1991fd,Sirlin:1991rt,Willenbrock:1991hu,Gegelia:1992kj,Gambino:1999ai
}.
   In contrast, defining the mass and width in terms of the complex-valued
position of the pole of the propagator leads to both
field-redefinition and gauge-parameter independence.

 As the baryon resonances are thought to be described by QCD, with the progress
of lattice techniques and, especially, the low-energy effective
theories (EFT) of QCD (see,
e.g., \cite{Weinberg:1979kz,Gasser:1983yg,Gasser:1988rb,Scherer:2005ri,%
Hemmert:1997ye,Hacker:2005fh,Pascalutsa:2006up} and references
therein) the question of defining baryon resonance masses becomes
important.
   Here we examine this issue for the $\Delta$ resonance.
   As discussed in Ref.~\cite{Hoehler}, the {\it
conventional resonance mass} and width determined from generalized
Breit-Wigner formulas have problems regarding their relation to
S-matrix theory and suffer from a strong model dependence.
   Here, we will show that these parameters, in addition, depend on the field-redefinition
parameter in a low-energy EFT of QCD.

   For simplicity we ignore isospin and consider an EFT of a single
nucleon, pion, and $\Delta$ resonance.
   Defining
$$
\Lambda_{\mu\nu}=-(i\,{\partial}\hspace{-.55em}/\hspace{.1em}
-m_\Delta) g_{\mu\nu}+i
\,(\gamma_{\mu}\partial_{\nu}+\gamma_{\nu}\partial_{\mu}) - i
\gamma_{\mu}{\partial}\hspace{-.55em}/\hspace{.1em}\gamma_{\nu}
-m_\Delta
\gamma_{\mu}\gamma_{\nu}, 
$$
   the free Lagrangian is given by
\begin{equation}
\mathcal{L}_{0}=
\bar{\psi}^{\mu}\,\Lambda_{\mu\nu}\,\psi^{\nu}+\bar\Psi(i\,{\partial}\hspace{-.55em}/\hspace{.1em}-m_N)\Psi
+\frac{1}{2}\,\partial_\mu\pi \partial^\mu\pi \,. \label{LfreiA}
\end{equation}
   Here, the vector-spinor $\psi^{\mu}$  describes
the $\Delta$ in the Rarita-Schwinger formalism
\cite{Rarita:1941mf},
   $\Psi$ stands for the nucleon field with mass $m_N$, and $\pi$
represents the pion field which we take massless to simplify the
calculations. The interaction terms have the form
\begin{equation}
\mathcal{L}_{\rm int}= g\, \partial^\nu \pi\, \bar\psi^\mu \left(
g_{\mu\nu}-\gamma_\mu\gamma_\nu \right)\Psi +
\mbox{H.c.}+\cdots\,, \label{Intlagrangian}
\end{equation}
where the ellipsis refers to an infinite number of interaction terms
which are present in the EFT. These terms also include all
counter-terms which take care of divergences appearing in our
calculations. Although our results are renormalization scheme
independent, for simplicity we use the dimensional regularization
with the minimal subtraction scheme.

Let us consider the field transformation
\begin{equation}
\bar\psi^\mu\to \bar\psi^\mu + \xi\,\partial^\mu\pi
\bar\Psi\,,\quad \psi^\nu\to \psi^\nu+\xi \,\partial^\nu\pi
\Psi\,,  \label{ftr}
\end{equation}
where $\xi$ is an arbitrary real parameter.
   When inserted into the Lagrangians of Eqs.~(\ref{LfreiA}) and
(\ref{Intlagrangian}), the field redefinition generates additional
interaction terms.
   The terms linear in $\xi$ are given by
\begin{equation}
\mathcal{L}_{\rm add\, int}= \xi\,\partial^\mu\pi\,
\bar\Psi\,\Lambda_{\mu\nu}\,\psi^{\nu}+ \xi \,
\partial^\nu\pi\, \bar{\psi}^{\mu}\,\Lambda_{\mu\nu}\,\Psi 
\,. \label{NewIntTerms}
\end{equation}
   Because of the equivalence theorem
physical quantities cannot depend on the field redefinition
parameter $\xi$.
   The complex-valued position of the pole of the $\Delta$
propagator does not depend on $\xi$. In contrast, the mass and
width defined via (the zero of) the real and imaginary parts of
the inverse propagator depend on $\xi$ at two-loop order.

The dressed propagator of the $\Delta$ in $n$ space-time dimensions
can be written as
\begin{eqnarray}
&-&i \left[ g^{\mu\nu}-\frac{\gamma^\mu\gamma^\nu}{n-1} -\frac{
p^\mu\gamma^\nu-\gamma^\mu p^\nu}{(n-1) m_{\Delta}}-\frac{(n-2)
p^\mu p^\nu }{(n-1) m_{\Delta}^2}\right]\nonumber\\
&& \times\frac{1}{p\hspace{-.35 em}/\hspace{.1em}-m_\Delta
-\Sigma_1 - p\hspace{-.4 em}/\hspace{.1em}\Sigma_6} +
\mbox{pole-free terms}\,,\label{dressedDpr}
\end{eqnarray}
where we parameterize the self-energy of the $\Delta$ as
\begin{eqnarray}
&& \Sigma_1(p^2) g^{\mu\nu}+\Sigma_2(p^2)
\gamma^{\mu}\gamma^{\nu}+\Sigma_3(p^2) p^{\mu}\gamma^{\nu}
+\Sigma_4(p^2) \gamma^{\mu}p^{\nu}\nonumber\\
&&+\Sigma_5(p^2)\,p^{\mu}p^{\nu} + \Sigma_6(p^2)\,
p\hspace{-.45em}/\hspace{.1em}
g^{\mu\nu}+\Sigma_7(p^2)\,p\hspace{-.45em}/\hspace{.1em}
\gamma^{\mu}\gamma^{\nu}\nonumber\\
&& +\Sigma_8(p^2)\, p\hspace{-.45em}/\hspace{.1em}
p^{\mu}\gamma^{\nu}+\Sigma_9(p^2)\, p\hspace{-.45em}/\hspace{.1em}
\gamma^{\mu}p^{\nu}+\Sigma_{10}(p^2)\,p\hspace{-.45em}/\hspace{.1em}
p^\mu p^\nu. \label{DseParametrization}
\end{eqnarray}
   The complex pole $z$ of the $\Delta$ propagator is obtained by solving the
equation
\begin{equation}
z - m_\Delta -\Sigma_1(z^2)-z\, \Sigma_6(z^2)=0\,.
\label{poleequation}
\end{equation}
The pole mass is defined as the real part of $z$.

On the other hand, the mass $m_R$ and width $\Gamma$ of the
$\Delta$ resonance are often determined from the real and
imaginary parts of the inverse propagator (corresponding to the
Breit-Wigner parametrization), i.e.,
\begin{eqnarray}
&& m_R - m_\Delta -{\rm Re}\, \Sigma_1(m_R^2)-m_R \, {\rm Re}\,
\Sigma_6(m_R^2)=0\,,\nonumber\\
&&\Gamma = -2\, {\rm Im}\, \Sigma_1(m_R^2)-2\, m_R \, {\rm Im}\,
\Sigma_6(m_R^2) \,. \label{Rmassequation}
\end{eqnarray}
   We have calculated the $\Delta$ mass using both definitions and analyzed their
$\xi$ dependence to first order (for details see
Ref.~\cite{Djukanovic:2007bw}).

The $\Delta$ self-energy at one loop-order is given by the diagram
in Fig.~\ref{DeltaMassInd:fig} (a). The two-loop contributions to
the $\Delta$ self-energy are given in Fig.~\ref{DeltaMassInd:fig}
(b) - (d). We are interested in terms linear in $\xi$.

To find the pole of the propagator we insert its loop expansion
\begin{equation}
z=m_\Delta+\delta_1 z+\delta_2 z+\cdots\, \label{polesparametr}
\end{equation}
in Eq.~(\ref{poleequation}) and solve the resulting equation order
by order.

The one-loop diagram results in the $\xi$-independent expression for
$\delta_1 z$. Calculating diagram (b) and (c) we find that they give
vanishing contributions. The $\xi$-dependent contributions in
$\delta_2 z$, generated by the one-loop diagram and by diagram (d)
exactly cancel each other leading to the $\xi$-independent pole of
the propagator.

\begin{figure}
\begin{center}
\epsfig{file=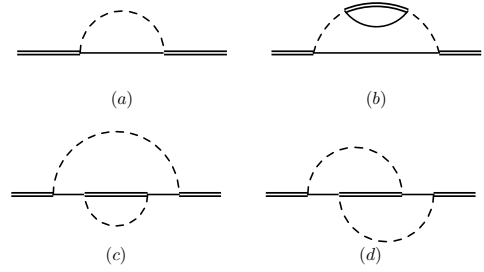,width=0.35\textwidth}
\caption[]{\label{DeltaMassInd:fig} $\Delta$ self-energy diagrams.
Solid, dashed, and double lines correspond to nucleon, pion, and
$\Delta$, respectively.}
\end{center}
\end{figure}

We perform the same analysis inserting the loop expansion of
$m_R$,
\begin{equation}
m_R=m_\Delta+\delta_1 m+\delta_2 m + \cdots,
\label{Rmassparametr}
\end{equation}
in Eq.~(\ref{Rmassequation}). For $\delta m_1$ the one-loop diagram
gives a $\xi$-independent expression. On the contrary, the
$\xi$-dependent contributions in $\delta_2 m$, generated by the
one-loop diagram and by diagram (d) do not cancel, thus leading to a
$\xi$-dependent mass $m_R$. An analogous result holds for the width
$\Gamma$ obtained from Eq.~(\ref{Rmassequation}).

\medskip

To conclude, we addressed the issue of defining the mass and width
of the $\Delta$ resonance in the framework of a low-energy EFT of
QCD. In general, the scattering amplitude of a resonant channel
can be presented as a sum of the resonant contribution expressed
in terms of the dressed propagator of the resonance and the
background contribution. The resonant contribution defines the
Breit-Wigner parameters through the real and imaginary parts of
the inverse (dressed) propagator. The resonant part and the
background separately depend on the chosen field variables, while
the sum is of course independent of this choice. We have performed
a particular field transformation with an arbitrary parameter
$\xi$ in the effective Lagrangian. In a two-loop calculation we
have demonstrated that the mass and width of the $\Delta$
resonance determined from the real and imaginary parts of the
inverse propagator depend on the choice of field variables. On the
other hand, the complex pole of the full propagator does not
depend on the field transformation parameter $\xi$.

The above conclusions are not restricted to the considered toy model
or EFT in general. Rather, our results are a manifestation of the
general feature that the (relativistic) Breit-Wigner parametrization
leads to model- and process-dependent masses and widths of
resonances. Although in some cases (like the $\Delta$ resonance) the
background has small numerical effect on the Breit-Wigner mass,
still the pole mass and the width should be considered preferable as
these are free of conceptual ambiguities.

\section{Bare and dressed quantities within a well defined model}
\label{sec:6}
\subsection{Longitudinal and transverse helicity amplitudes of nucleon resonances
in a constituent quark model - bare vs dressed resonance couplings}
Many models have been built and applied to the description
of hadron properties. An important role is played by Constituent
Quark Models (CQM), in which quarks are considered as
effective degrees of freedom. There are many versions of
 CQM, which differ according to the chosen quark dynamics:
h.o. and three-body force \cite{ho}, algebraic \cite{bil},
hypercentral \cite{pl}, Goldstone Boson Exchange \cite{olof},
instanton \cite{bn}. Here we report some results of the hypercentral
CQM (hCQM) \cite{pl} on the longitudinal and transverse helicity
amplitudes of the nucleon resonances. In this model, the quark
interaction is assumed to be given by a hypercentral potential
\begin{equation}  \label{eq:hpot}
V(x)~=~-\tau /x~ +~ \alpha~ x, ~~~~~~~~~x~ =~\sqrt{\rho^2 + \lambda^2}
\end{equation}
where $x$ is
the hyperradius expressed in terms of the internal Jacobi coordinates
$\vec{\rho}$ and $\vec{\lambda}$.
A coulomb-like plus linear confinement form of the potential is supported
by recent lattice QCD evaluations
of the quark-antiquark potential \cite{bali} and in this sense
Eq. (\ref{eq:hpot}) can be considered as the hypercentral
approximation of a two-body Cornell-like potential. The model interaction is
completed by adding a standard spin dependent hyperfine interaction
$H_{hyp}$ \cite{ho}, in order to reproduce the splittings within the
$SU(6)$ multiplets. The few free parameters ($\alpha, \tau$ and the
strength of $H_{hyp}$) are fitted to the spectrum and the model is
then applied to calculate (i.e. to predict) various properties of
hadrons: the photocouplings \cite{aie}, the transverse helicity
amplitudes for negative parity resonances \cite{aie2}, the elastic
form factors \cite{el}, the longitudinal and transverse helicity
amplitudes of all the main resonances \cite{ms}.

\begin{figure}[ht]
\begin{center}
\resizebox{0.4\textwidth}{!}{%
  \includegraphics{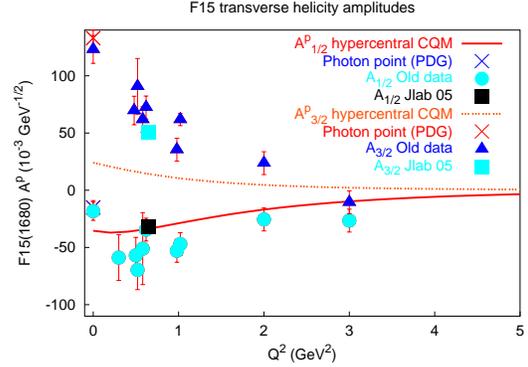}
}
\caption{The transverse helicity amplitudes for the $F_{15}$ resonance
obtained with the hCQM \cite{ms}, compared with the experimental
data,
taken by an old compilation \cite{bur}, recent JLab experiments
\cite{jlab05} and PDG \cite{pdg}. }
\label{f15}
\end{center}
\end{figure}

It is interesting to analyze in a systematic way the $Q^2$ behaviour of
the helicity amplitudes
in comparison with the existing data. Fig. \ref{f15} shows the results for the
transverse helicity amplitudes of the $F_{15}$ resonance, results which are
typical for a $J>1/2$ state \cite{aie2,ms}: the
medium-high $Q^2$ behaviour is quite well reproduced, showing that the
hypercoulomb part of the interaction $1/x$ gives a fair account of the
short range, while at low $Q^2$ there is a lack of strength,
particularly for the $A_{3/2}$ amplitude.
For $J=1/2$
states \cite{aie,ms}, there are some minor problems in the low
$Q^2$ region, but for the rest the agreement with data is satisfactory.
Major problems are present for the Roper resonance \cite{ms}, a fact that
may support the idea of a particular status of the radial excitations
of the nucleon. Discrepancies are present also for the $\Delta -$resonance
\cite{ms}, a feature which is typical of all CQMs; it is well known that
the quark model, while reproducing quite well the baryon magnetic moments,
fails in the case of
the $N-\Delta$ transition magnetic moment.
Taking into account the fact that the proton radius, calculated with the
wave functions corresponding to the potential of Eq. (\ref{eq:hpot}),
turns out to be about $0.5 fm$, the emerging
picture is that of a small quark core surrounded by an external region,
which is probably dominated
by dynamical effects not present in the CQM, that is sea-quark or meson
cloud effects \cite{aie2}.

These considerations are relevant in connection with the issue of
bare vs. dressed quantities. One should not forget that the
separation  between bare and dressed quantities is meaningful within
a definite theoretical approach. In CQM calculations the aim is not
a fit but the description of observables, which in principle are
dressed quantities (like baryon masses, magnetic moments, helicity
amplitudes, etc).  In any case the identification of quark results
with bare quantities is questionable in view of the fact that CQs
have a mass and some dressing is implicitly taken into account.
However a consistent and systematic CQM approach may be helpful in
order to put in evidence explicit dressing effects.

These effects have been recently
calculated by means of a dynamical model \cite{dmt}. The meson cloud
contribution to
various helicity amplitudes has been calculated and compared with the hCQM
predictions
\cite{ts03}. The two contributions cannot be added, since they are
calculated within
different frameworks, however it is interesting to note that the meson
cloud
contribution is relevant at low $Q^2$ and in most cases it is important
where the
hCQM prediction underestimates the data. An example of this situation
is given by the longitudinal and transverse helicity amplitudes for the
$\Delta -$excitation \cite{ts03}. The case of the $S_{1/2}$ amplitude is
particularly interesting (see Fig. \ref{delta}) : the hCQM is
almost vanishing and the meson cloud contribution accounts for practically
the whole strength.

\begin{figure}[htb]
\begin{center}
\includegraphics{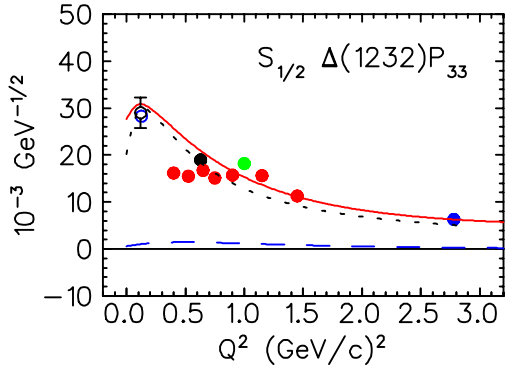}
\caption{ The $Q^2$ dependence of the $N \rightarrow \Delta$
longitudinal helicity
amplitude $S_{1/2}$: superglobal fit performed with MAID \cite{maid}
(solid curve),
predictions of the hypercentral Constituent Quark Model
\cite{pl,ts03,ms} (dashed curve), pion cloud contributions calculated
with the Mainz dynamical model \cite{dmt} (dotted curve). The data points
at finite $Q^2$ are the results of single-Q$^2$ fits \cite{ts03} on recent
data quoted in ref. \cite{ts03}. }
\label{delta}
\end{center}
\end{figure}

The
problem is how to introduce dressing in the calculations.
One way is to adopt a
hadronic approach: meson and baryons (nucleon and nucleon resonances) are
the relevant
degrees of freedom and the dynamics is given by meson-baryon interactions.
This is
certainly a consistent approach which has been used with success by
various groups with
different techniques (see e.g. \cite{maid,mz,anl}). Another
possibility is given by
the so called hybrid models, where the baryons are considered as
three-quark states surrounded by a pion cloud
and a direct quark-meson coupling is introduced. In this way the
electromagnetic
excitation acquires  contributions also from the meson cloud.
This
approach is very useful for preliminary calculations (see ref.
\cite{dong}), however a
more promising method is provided by the inclusion of dressing mechanisms
directly at
the quark level. This means in particular the inclusion of higher Fock
components in the
baryon state:
\begin{equation}
|\Psi_{B}>~=~ \Psi_{3q}~|qqq>~+~\Psi_{3q~q \bar{q}}~|qqq~q \bar{q}>
\end{equation}
and implies the necessity of unquenching the quark model, as discussed in
section $2.1$. For the case of mesons, there are pioneering works by
Geiger and Isgur
\cite{gi}, where the q\={q} pair creation mechanism is introduced at the
microscopic
level within a string model. In the case of baryons, the
problem is more complicated and has been recently treated performing the
sum over intermediate quark loops by means of group
theoretic methods \cite{sb}. This approach has been applied to the
determination
of the strange content of the proton \cite{bs} with good results. In this way we
have at our disposal a promising method for obtaining an unquenched,
that is dressed,
formulation of the CQM. The systematic calculation of baryon properties, such as
transition amplitudes (but also elastic form factors and
structure functions) in an unquenched CQM will supply a set of dressed
quantities to be
compared directly with data and will allow to
understand where meson cloud or (better) q\={q} effects are important.

One should however be aware of some problems, both phenomenological and
theoretical. From the phenomenological point of view, there is the problem
of the sign of the helicity amplitudes, which is actually extracted from
the meson electroproduction amplitude, the fact that the PDG photon points
are often non consistent and the need of new and systematic data. The main
theoretical problems are connected with the inclusion of relativity. The
kinematic relativistic corrections seem to be not important for the helicity
amplitudes \cite{mds2}, however relativity should be consistently included
both in the (unquenched) CQM states and in the transition operators,
leading to the possibility of quark pair terms in the electromagnetic
current.
In any case, the unquenching of the CQM is expected to produce a substantial
improvement in the theoretical description of baryon properties. In
particular, it will be possible to calculate simultaneously
the electromagnetic processes and the strong decays and the baryons
resonances will acquire a non zero width through the coupling to the
continuum part of the spectrum.

\subsection{Nucleon Resonances and Hadron Structure Calculations}
\subsubsection{What are the nucleon resonances ?}

To answer this question, it is useful to recall some
textbook (For example, see Refs.\cite{gw,bm,feshbach}) definition of
resonances. Phenomenologically, a resonance ($R$) is identified with a peak in
a plot of the reaction cross section as function of the
collision energy $E$ or invariant mass $W$. At energies near the peak position,
one can fit such data $near$ the peak by
\begin{eqnarray}
\sigma_{a,b} (W\sim m_R) &\sim & \rho(W)\frac{|\Gamma_{R,a}|^2| \Gamma_{R,b}|^2}
{(W- m_R)^2 +|\frac{\Gamma_0}{2}|^2}
\label{eq:rcs}
\end{eqnarray}
where $\rho(W)$ is an appropriate phase space factor,
$m_R$ the position of the peak, and $\Gamma_0$ the width of the
peak.
The expression Eq.(\ref{eq:rcs}) has the same function form
of the decay rate of an unstable system with a mass $\sim m_R$ and a
life time $\tau_R \sim 1/\Gamma_0$. It is thus natural to interpret that the cross section
Eq.(\ref{eq:rcs}) is due to the excitation of an unstable system during
the collision.
How this unstable system is formed from the entrance and
exit channels is a dynamical question which can only be answered by
modeling the reaction mechanisms and the internal structure of all particles
involved.
We will address this question within the Hamiltonian formulation of
the problem. This is rather different from the S-matrix approach.

Within the Hamiltonian formulation, there are two ways to derive the
expression Eq.(\ref{eq:rcs}) depending on the structure of the excited unstable
system. Let us first consider the one defined in Feshbach's textbook
(page 23 of Ref.\cite{feshbach}). It can be stated as the following :

\vspace{0.5cm}
{\bf A resonance is formed in a process that
 the incident projectile completely lose its identity, amalgamating
with the target system to form a compound state. Namely, the
evolution of the whole system can not be defined in terms of the motion
of the projectile and its transmutation. }
\vspace{0,5cm}

The expression Eq.(\ref{eq:rcs}) corresponding to this definition of
resonances can be formulated by assuming that the Hamiltonian of the system
has the following form
\begin{eqnarray}
H=H_0 + H'
\label{eq:isobarh0}
\end{eqnarray}
with
\begin{eqnarray}
H' = \sum_{a} \Gamma_{R,a}
\label{eq:isobarh}
\end{eqnarray}
where $\Gamma_{R,a}$ defines the decay of an unstable system $R$ with
a mass $M_0$ into
channel $a$.
The reaction amplitude is defined by
\begin{eqnarray}
T(E)= H' + H'\frac{1}{E-H+i\epsilon}H'
\label{eq:tmatrix}
\end{eqnarray}
>From Eqs.(\ref{eq:isobarh})-(\ref{eq:tmatrix}), it is
straightforward to see that the reaction cross section for $b\rightarrow a$
can be written as
\begin{eqnarray}
\sigma_{a,b} (W) &= &\rho(W) |T^R_{a,b}(W)|
\label{eq:rcs-1}
\end{eqnarray}
with
\begin{eqnarray}
T^R_{a,b}(W)= \frac{\Gamma_{R,a}(k_a) \Gamma_{R,b}(k_b)}
{W- M_0-\Sigma(W)}
\label{eq:tr}
 \end{eqnarray}
where $k_a$ is the on-shell momentum of channel $a$ and
\begin{eqnarray}
\Sigma_R(W)=
\sum_{a}<R|\Gamma^\dagger_{R,a}
\frac{1}{W-H_0+i\epsilon} \Gamma_{R,a}|R>
\label{eq:mass-shift}
\end{eqnarray}
We can cast Eq.(\ref{eq:tr}) into
\begin{eqnarray}
T^R_{a,b}(W)=
\frac{\Gamma^*_{R,a}(k_a) \Gamma_{R,b}(k_b)}
{W- M_R(W) +i\frac{\Gamma^{tot}(W)}{2}}
\label{eq:tbw}
\end{eqnarray}
where
\begin{eqnarray}
M_R(W) &=& M^0+ Re(\Sigma_R(W))
\label{eq:isobar-pole-1} \\
\frac{\Gamma^{tot}(W)}{2} &=& - Im (\Sigma_R(W))
\label{eq:isobar-pole-2}
\end{eqnarray}

By using Eqs.(\ref{eq:rcs-1}) and (\ref{eq:tbw}) to fit the
expression Eq.(\ref{eq:rcs}), the  parameters of the Hamiltonian are
then related to the data by the following relations
\begin{eqnarray}
m_R&=& M_R(W=m_R) = M^0+ Re(\Sigma_R(m_R))
\label{eq:bw-pole-1}\\
\Gamma_0&=&{\Gamma^{tot}(W=m_R)} =-2 Im (\Sigma_R(m_R))
\label{eq:bw-pole-2}
\end{eqnarray}
Eqs.(\ref{eq:bw-pole-1})-(\ref{eq:bw-pole-2}) then allow us
to use the experimental
values $m_R$ and $\Gamma_0$ to extract the property of the unstable system,
specified by $M_0$ and $\Gamma_{R,a}$ of the Hamiltonian, through the
evaluation of
Eqs.(\ref{eq:mass-shift})-(\ref{eq:isobar-pole-1})
 at energies near $W=m_R$.

The second mechanism which can also yield a
cross section of the form of Eq.(\ref{eq:rcs}) is :

\vspace{0.5cm}
{\bf An unstable system is formed during the collision
by an attractive force between
the interacting particles which do not lose their
identities. }

\vspace{0.5cm}
The simplest parameterization of an attractive force is a separable potential
\begin{eqnarray}
H^\prime = g^\dagger\frac{1}{C}g
\end{eqnarray}
The solution of Eq.(\ref{eq:tmatrix}) then become
\begin{eqnarray}
T(W)=\frac{g^*(k)g(k)}{C-z(W)}
\label{eq:tpot}
\end{eqnarray}
where
\begin{eqnarray}
z(W)=<g|\frac{1}{W-H_0+i\epsilon}| g>
\end{eqnarray}
If the parameters of $H^\prime$ are chosen such that
$C-Re(z(W)) \rightarrow 0 $ on the physical world $W \rightarrow W_0$ where $W_0$
is a real number, we can expand
\begin{eqnarray}
& C&-z(W) =  \\
&[&C- R(W_0)-R^\prime(W_0)(W-W_0)+\cdot\cdot\cdot] - i I(W)  \nonumber \\
&\sim& - R^\prime(W_0) \nonumber  \\
& [& W-W_0 - \frac{1}{R^\prime(W_0)}(R(W_0)-C) \nonumber
+i\frac{1}{R^\prime(W_0)}I(W)] \nonumber
\end{eqnarray}
where $R(W_0) = Re(z(W_0))$, $I(W)=Im(z(W))$, and \\
$R^\prime (W_0) = \partial Re(z(W))/\partial W|_{W=W_0}$.
We then can write at $W \rightarrow W_0$
\begin{eqnarray}
& T& (W\sim W_0 )= \\
& & \frac{-g^*_a(k_0)\frac{1}{R^\prime(W_0)}g_b(k_0)}
{W-[W_0+\frac{1}{R^\prime(W_0)}(R(W_0)-C)]+i\frac{1}{R^\prime(W_0)}I(W_0)} \nonumber
\end{eqnarray}
The above expression can give resonant cross section Eq.(\ref{eq:rcs}) if the
parameters of $g$ and $C$ can be chosen to satisfy
\begin{eqnarray}
m_R &=& W_0+\frac{1}{R^\prime(W_0)}(R(W_0)-C) \\
\frac{\Gamma_0}{2} &=&  \frac{1}{R^\prime(W_0)}I(W_0)
\end{eqnarray}

\subsubsection{Dynamical Models for investigating
Nucleon Resonances}
>From the above two examples, we see  that the resonant cross section
Eq.(\ref{eq:rcs}) can correspond to two very different internal structure of
the excited unstable system. The nucleon resonances we are interested in
correspond to the
unstable systems defined by the Hamiltonian Eq.(\ref{eq:isobarh}). For the
meson-nucleon reactions, such unstable systems are due to the excitation of
the quark-gluon substructure of the nucleon.

In reality, the situation is much more complicated. In the reactions
involving composite systems, such as atoms, nuclei and nucleons, the
excitations of resonances always involve non-resonant
direct interactions. For example, the non-resonant interactions
in pion-nucleon scattering could be due to the exchange of $\rho$
meson. The reaction formulation for analyzing such reactions is well
presented in Feshbach's textbook\cite{feshbach}. We now briefly
describe how such a formulation can be used to investigate
nucleon resonances in meson-nucleon reactions.

The starting point is to divide the Hilbert space into a $P$ space
for the entrance and exit channels
 and $Q$ for the rest.  One can cast the equation of motion in the
$P$-space as
\begin{eqnarray}
(E-H_{eff})P\Psi =0,
\end{eqnarray}
where
\begin{eqnarray}
H_{eff}=H_{PP}+H_{PQ}\frac{1}{E^{(+)}-H_{QQ}}H_{QP}\,.
\end{eqnarray}
Here $E^{(+)} = E + i\epsilon$ specifies the boundary condition and
we have defined projected operator $H_{PP}=PHP$, $H_{PQ}=PHQ$ and
$H_{QQ}=QHQ$. Now consider the eigenstates of $H_{QQ}$ which can be
discrete bound $\Phi_{s}$ or unbound $\Phi_{\epsilon,\alpha}$ states
\begin{eqnarray}
H_{QQ}\Phi_{s}&=&\epsilon_s \Phi_{s}
\label{eq:hqqb}\\
H_{QQ}\Phi(\epsilon,\alpha)&=&\epsilon \Phi(\epsilon,\alpha)
\label{eq:hqqc}
\end{eqnarray}
with
\begin{eqnarray}
<\Phi_{s}|\Phi_{s'}>&=&\delta_{s,s'}
\\
<\Phi(\epsilon,\alpha)|\Phi(\epsilon',\alpha')>&=&\delta_{\alpha,\alpha'}
\delta(\epsilon-\epsilon')
\end{eqnarray}

We then expand
\begin{eqnarray}
&H_{eff}&-H_{PP}= \nonumber \\
& & \sum_{s}\frac{H_{PQ}\Phi_s><\Phi_s H_{QP}}{E-\epsilon_s} + \nonumber \\
&+ & \int d\alpha \int d\epsilon  \frac{H_{PQ}\Phi(\epsilon,\alpha)><\Phi(\epsilon,\alpha) H_{QP}}
{E^{(+)}-\epsilon}
\end{eqnarray}
One can see from the above equation that rapid energy dependence of
$H_{eff}$ will occur as the energy approaches one of the bound state energy
$\epsilon_s$. This is the origin of rapid energy-dependence of the cross
sections. As shown in Feshbach's book (page 158-162), the amplitude at
$E\sim \epsilon_s$ can be written as
\begin{eqnarray}
T_{fi}&=&T^P_{fi}+ \nonumber \\
&+& \frac{<\chi^{(-)}|H_{PQ}|\Phi_s><\Phi_s|H_{QP}\chi^{(+)}>}
{E-\epsilon_s-<\Phi_s|W_{QQ}|\Phi_s>}
\label{eq:fesh-t}
\end{eqnarray}
with
\begin{eqnarray}
W_{QQ} = H_{QP}\frac{1}{E^{(+)}-\hat{H}_{PP}}H_{PQ}
\label{eq:fesh-selfe}
\end{eqnarray}
where
\begin{eqnarray}
&\hat{H}_{PP}& =H_{PP}+ \nonumber \\
& & \int d\alpha \int d\epsilon
\frac{H_{PQ}\Phi(\epsilon,\alpha)><\Phi(\epsilon,\alpha) H_{QP}}
{E^{(+)}-\epsilon}
\label{eq:hhatpp}
\end{eqnarray}
and $\chi^{(\pm)}$ are the solutions of
\begin{eqnarray}
& &(E-\hat{H}_{PP})\chi^{(+)}=0
\label{eq:fesh-chip}\\
& &(E-\hat{H}_{PP})\chi^{(-)*}=0
\label{eq:fesh-chim}
\end{eqnarray}
In this formulation, one bound state of $H_{QQ}$ will correspond to
one resonance. Namely, one can predict whether a resonance can appear in a
particular partition of Hilbert space by examining whether
bound states can be generated from the Hamiltonian when the
coupling with the states $P$ are turned off.

We now point out that the dynamical model developed in
Ref.\cite{Sat} (MSL model)
for investigating meson-baryon reactions are completely consistent with
the formulation given in Eqs.(\ref{eq:fesh-t})-(\ref{eq:fesh-chim}).
To see this, one just make the following identifications:
\begin{itemize}
\item $P$ space contains reaction channels \\ \mbox{$MB = \pi N, \gamma N, \eta N,
\pi\Delta, \rho N, \sigma N$ and $\pi\pi N$}
\begin{eqnarray}
P = \sum_{MB} |MB><MB| + |\pi\pi N><\pi\pi N|
\end{eqnarray}
\item $H_{QQ}$ describes the internal structure of the bare $N^*$
states
\begin{eqnarray}
H_{QQ}|N^*_i> &=& M^0_{N^*_i}|N^*_i>
\label{eq:hqq} \nonumber \\
Q &=& \sum_{i} |N^*_i><N^*_i|
\end{eqnarray}
\item $H_{PP}$ defines  the non-resonant meson-baryon interactions
 \begin{eqnarray}
&H_{PP}&= \sum_{MB} |MB>[\sqrt{m_B+p^2}+\sqrt{m_B+\vec{k}^2}]<MB|  \nonumber \\
&+& \sum_{MB,M'B'}v_{MB,M'B'}
+\sum_{MB}[v_{MB,\pi\pi N} + v_{\pi\pi N,MB}] \nonumber \\
 &+&  v_{\pi\pi N,\pi\pi N}
\end{eqnarray}
\item $H_{QP}$ defines the coupling of the internal structure of $N^*$ with
the reaction channels
\begin{eqnarray}
H_{QP}= \sum_{N^*}[(\sum_{MB}\Gamma_{N^*,MB} + \Gamma_{N^*,\pi\pi N}]
\end{eqnarray}
\end{itemize}

With some inspections, one can see that equations
presented in the section 3 of Ref.\cite{Sat} (MSL model)
are completely equivalent to Eqs.(\ref{eq:fesh-t})-(\ref{eq:fesh-chim}).
If we set $MB=\pi N, \gamma N$ and $Q = |\Delta><\Delta|$, and
neglect $\pi\pi N$ channel,
we then obtain the formulation of the SL model
\cite{sl}.

\subsubsection{Relations with Hadron Structure Calculations}
We now note that $\epsilon_s$ and $\Phi_s$ in Eq.(\ref{eq:hqqb})
and (\ref{eq:fesh-t}) relate
the structure calculations for the unstable systems in the $Q$-space
and the reaction amplitudes
in the P-space. In the MSL formulation, these are the
bare mass $M^0_{N^*_i}$ and wave function $|N^*>$ of
the discrete $bound$ states defined by Eq.(\ref{eq:hqq}).
These bare $N^*$ states can be considered as
the excited states of
the nucleon if its coupling with the reaction channels is turned off.
It is therefore
natural to speculate that the bare $N^*$ states in the SL and MSL models
 correspond
to the predictions from a hadron model with confinement force, such
as the well-developed constituent quark model with gluon-exchange interactions.
This was first noticed in the SL model in 1996. The idea was later pursued
in Ref.\cite{yoshimoto} in 2000 in an attempt to directly calculate $\pi N$
scattering amplitude in $S_{11}$ up to $W=2$ GeV starting from several
constituent quark models. In the later $N(e,e^\prime N)$
analysis\cite{sl1,jlss} based on SL
model, the bare $\gamma N \rightarrow \Delta$ (1232) form factors  were
 also found to
be close to the constituent quark model predictions. To consider
constituent quark models with meson-exchange residual interactions,
the SL and MSL models must be modified to account for the
contribution due to the continuum in the $Q$-space; namely the
effects due to the second term in the right-hand-side of
Eq.(\ref{eq:hhatpp}).

It is unlikely that the Lattice QCD calculation (LQCD)
can account for the channel coupling effects and unitarity conditions,
which are the essential elements of a dynamical coupled-channel
analysis,
rigorously in the near future. It is a challenging problem
to relate the LQCD
calculations to the information which can be
extracted from the full solution
Eq.(\ref{eq:fesh-t}) of a dynamical coupled-channel analysis.

One possibility is to perform a LQCD calculation which defines a
$H_{QQ}$ of a dynamical coupled-channel analysis. At the present
time, perhaps the predictions from a quenched LQCD with heavy quark
mass and $no$ chiral extrapolation correspond to the bare parameters
resulted from the dynamical coupled-channel analysis being performed
at EBAC. This is based on the argument that the quark-loop
contributions are suppressed at heavy quark limit and the LQCD
mainly accounts for the gluonic interactions which are not in the
$P$-space of MSL model.

\section{Conclusion}
This work was motivated by problems relating the reliable results from
partial-wave and amplitude analysis, which are the parameters of
dressed scattering matrix singularities, and the results of quark
models, which are usually given as the properties of bare resonances.
Undressing dressed scattering matrix singularities in coupled-channel
models involves model-dependent hadronic mass shifts, which arise from
the unmeasurability of off-shell effects accompanying the dressing
procedure. It is legitimate to extract bare quantities in
coupled-channel models within the framework of a well defined model,
but their interpretation requires keeping track of hadronic mass shifts
produced by off-shell ambiguities. The best meeting point between quark
model predictions and analyses of experimental data are dressed
scattering matrix singularities, as although dressing (un-quenching)
the quark model is complicated, it is in principle a solvable problem.
This will require careful definition and checking of the procedures for
extracting poles from energy-dependent partial waves or directly from
partial-wave data.

%
%

\end{document}